\documentclass[journal,twocolumn]{IEEEtran}

\usepackage{color}

\usepackage[font=footnotesize]{subfig}

\ifCLASSINFOpdf
\usepackage[pdftex]{graphicx}
\graphicspath{{images/}}
\else
\fi

\usepackage[cmex10]{amsmath}
\graphicspath{{images/}}
\usepackage{parskip} 
\usepackage{textcomp} 
\usepackage{amssymb} 
\usepackage{enumerate}
\usepackage{array}
\usepackage{fancyhdr}
\usepackage{hyperref}
\usepackage{xfrac}    
\usepackage[ugly]{nicefrac}

\def\D{{\mathbf D}}

\def\H{{\mathbf H}}

\def\0{{\mathbf 0}}
\def\1{{\mathbf 1}}

\def\a{{\mathbf a}}
\def\b{{\mathbf b}}

\def\g{{\mathbf g}}
\def\h{{\mathbf h}}

\def\t{{\mathbf t}}

\def\w{{\mathbf w}}
\def\x{{\mathbf x}}
\def\y{{\mathbf y}}

\newcommand{\ex}[1]{\mathbb{E}\left[#1\right]}

\newcommand{\openone}{\leavevmode\hbox{\small1\normalsize\kern-.33em1}}

\usepackage[table,xcdraw]{xcolor}
\usepackage{multirow}
\usepackage{slashbox}

\begin{document}

\title{Neural Network Aided Computation of Mutual Information for Adaptation of Spatial Modulation}

\author{
\IEEEauthorblockN{Anxo Tato\IEEEauthorrefmark{1}, Carlos Mosquera\IEEEauthorrefmark{1}, Pol Henarejos\IEEEauthorrefmark{2}, Ana P\'erez-Neira\IEEEauthorrefmark{2}\IEEEauthorrefmark{3}} \\ \vspace{1em}
\IEEEauthorblockA{\IEEEauthorrefmark{1} atlanTTic Research Center, Universidade de Vigo, Galicia, Spain} \\
\IEEEauthorblockA{\IEEEauthorrefmark{2} Centre Tecnol\`ogic de Telecomunicacions de Catalunya (CTTC), Castelldefels, Spain} \\
\IEEEauthorblockA{\IEEEauthorrefmark{3} Dept. of Signal Theory and Communications, Universitat Polit\`ecnica de Catalunya (TSC-UPC)} \\
\IEEEauthorblockA{Email: \{anxotato, mosquera\}@gts.uvigo.es, \{pol.henarejos, ana.perez\}@cttc.cat}
}

\maketitle

\begin{abstract}
Index Modulations, in the form of Spatial Modulation or Polarized Modulation, are gaining traction for both satellite and  terrestrial next generation  communication systems.  Adaptive Spatial Modulation based links are needed to fully exploit 
the transmission capacity of time-variant channels. The adaptation of code and/or  modulation requires a real-time evaluation of the channel achievable rates. Some existing results in the literature present a  computational complexity which scales quadratically with the number of transmit antennas and the constellation order. Moreover, the accuracy of these approximations is low and it can lead to wrong Modulation and Coding Scheme selection. In this work we apply a  Multilayer Feedforward Neural Network to compute the achievable rate of a generic Index Modulation link. The case of two antennas/polarizations is analyzed throughly  showing the neural network not only a one-hundred fold decrement of the Mean Square Error in the estimation of the capacity compared with existing analytical approximations, but it also reduces fifty times the computational complexity. Moreover, the extension to an arbitrary number of antennas is explained and supported with simulations. More generally, neural networks can be considered as promising candidates for the practical estimation of complex metrics in communication related settings.
 \end{abstract}


\section{Introduction}

The evaluation of the achievable physical layer rate of   a given modulation scheme is an important theoretical problem with very relevant use in practice.  Most  modern communication standards, e.g.,  \cite{LTE-R14}, \cite{Wifi-2013} or \cite{S-UMTS-family-SL}, incorporate  some sort of Adaptive Coding and Modulation (ACM) mechanism, generically known as  link adaptation. This consists typically on varying the modulation order and/or the coding rate of the channel encoder to track the varying channel conditions. The ultimate goal is to adjust the transmitted bit rate to the information that the channel can support for a given bit error probability. 

Link adaptation makes it necessary for the transmitter to estimate somehow the  mutual information (MI) between the transmit and received waveforms on a per-frame basis, so that the most efficient  Modulation and Coding Scheme (MCS) can be chosen. In most cases, the receiver computes some metric related to the MI and sends it back to the transmitter end. This metric can be in the form of the average or effective Signal to Interference and Noise Ratio (SINR), or some Channel Quality Indicator (CQI) specifically suited to the set of MCS available to the transmitter \cite{Wimax-MMIB}. In essence, the receiver must estimate the maximum amount of information that can be transmitted reliably through the channel; for all this, the estimation of the MI plays an instrumental role. 

A particular family of modulation schemes, known as  Index Modulations (IM) \cite{Basar-2016-IM-5G}, have attracted a great deal of interest in the last few years.  Among others, we can cite Spatial Modulation (SM) \cite{Mesleh-2006}-\cite{Yang-2015-guidelines-SM} or Polarized Modulation (PMod) \cite{PMod-CTTC}. SM and its more sophisticated variants are proposed for next generation of wireless networks due to several advantages. In comparison to  single-antenna techniques, the spectral efficiency increases, with a simpler and more energy efficiency hardware than in other multi-antenna techniques \cite{Basar-2016-IM-5G}.   
Another interesting version is that studied in \cite{PMod-CTTC}, where the authors propose the use of PMod to increase the spectral efficiency of next generation mobile satellite communications; if Multiple-Input-Multiple-Output (MIMO) signal processing techniques are applied to Dual Polarization (DP) satellite systems, the performance of single-antenna (or single polarization) links can be notably enhanced. DP schemes were also highlighted in  \cite{DP-channel-Nikolaidis-2018} as a means to improve the satellite coverage in remote areas to serve the increasing number of Internet of Things (IoT) devices. 

In this paper we present a novel method to compute the mutual information without Channel State Information at the Transmitter (CSIT) of a 2$\times$2 SM system, and show how to generalize it to an arbitrary number of antennas. The results are also valid to other types of IM, like PMod. This calculation is needed, for example, in an adaptive SM system where the transmitter uses this mutual information, obtained and fed back by the receiver, to select the proper MCS. Results requiring numerical integration, for example by means of  Monte Carlo simulations,  can be found in the literature \cite{Narasimhan-2016-letter-GSM} and \cite{ICASSP-18-CTTC}. One value of this work is that it explores a radically new approach to solve an essential problem in the practical application of Information Theory:  the mutual information of non-conventional modulations is computed by means of Machine Learning (ML) tools.


The use of ML at the physical layer of communication systems is gaining momentum, see the recent surveys \cite{Ahad-2016-NN-Wireless} and \cite{Simeone-2018-ML-comms}. In particular,  Neural Networks (NN) have been successfully used for channel estimation and equalization \cite{Moustafa-2009-NN}, signal recognition and modulation classification \cite{Elsoufi-2016-NN}, \cite {Hanna-2017-NN}, detection  in MIMO Generalized SM \cite{Marseet-2017-MS-GSM}, and learning of physical layer parameters in Cognitive Radio \cite{Wyglinsnki-2017-CR-DNN}, among others. In \cite{Kassab-2009} and \cite{Sandanalakshmi-2013} NNs are applied to perform link adaptation in multicarrier systems. In \cite{Baldo-08-ICC-NN-Wifi}, a Multilayer Feedforward Neural Network (MFNN)  is used to predict the performance of a WiFi cell. A deep NN is proposed in 
\cite{Sun-SPAWC-17} to decide the optimal power allocation in a wireless resource management problem. In this latter reference the  NN is used to obtain the optimal power allocation values, much more efficiently (by speeding up the computational time in several orders of magnitude) than  the baseline iterative algorithm which solves the corresponding non-convex optimization problem. Besides NNs, Support Vector Machines (SVM) have been also studied for the selection  of physical layer parameters in communication settings with a large number of  degrees of freedom  \cite{Alberto-MIMO}.

The current work applies a one-hidden layer MFNN as a facilitator scheme to compute the MI in an adaptive $2\times 2$ SM link, based on some specifically selected input features which can be easily obtained from the MIMO channel matrix, together with  the  Signal to Noise Ratio (SNR). To the best of our knowledge, it is the first time that a NN is proposed to estimate the MI of a channel. In the particular scenario of SM, the evaluation of the capacity, needed for adaptation purposes, is numerically demanding when needed on-the-fly. 

The shallow NN proposed to calculate the MI of a generic SM system, valid also for PMod, outperforms recent approximations found in the literature such as \cite{ICASSP-18-CTTC} and \cite{Guo-2014-MI-SM-ICC}, both in terms of estimation accuracy  and computational complexity. In order to avoid the numerical evaluation of the involved integrals, these references provide two different approximations of the MI for a specific symbol constellation, with a complexity scaling  with the square of the constellation size and the number of antennas. As opposed, the proposed solution has a much lower complexity, which is independent of the size of the constellation. 

The main contribution of this work is the accurate evaluation of the MI of spatial modulations, which  have resisted so far those attempts to obtain simple expressions. Moderate size standard neural networks will be seen to be up to the task provided that a careful extraction of channel parameters and training are performed. 
The proposed solution to calculate the MI enables also to perform a different type of adaptation in SM systems. Due to the lack of accurate MI evaluation methods, works like \cite{Yang-2015-guidelines-SM} and \cite{Yang-2012-LA-SM} only consider the selection of the modulation order in an adaptive system.  However, with the accurate MI obtained with the NN, the transmitter could not only adapt the modulation scheme, but also the coding rate of the channel encoder, allowing a fine granularity in the available transmit MCS. Moreover, due to the lower complexity of the proposed method, the receiver can compute  the MI more often and then  follow  faster channel variations, so that the adaptation speed  is not necessarily limited by the complexity of the computation of  the channel capacity.  
 

The rest of the paper is structured as follows. Section \ref{sec:system-model} explains our system model and introduces the reader into Spatial Modulation. Then, Section \ref{sec:analytical} presents the expressions to compute the mutual information of SM.  It also replicates the analytical expressions existing in the literature to approximate the MI, and to be used for benchmarking purposes. Afterwards, in Section \ref{sec:MFNN} a brief introduction to Multilayer Feedforward Neural Networks is included before dealing with their specific  application to the evaluation of the  MI of a $2\times2$ SM for different constellations. Afterwards, Section \ref{sec:results} presents the simulation results in detail for the case of two dimensions. Then, Section \ref{sec:extension} explains how to generalize the method to obtain the MI of systems with a higher number of antennas. Lastly, the main conclusions are drawn in Section \ref{sec:conclusions}.

\noindent \textit{Notation:} Upper (lower) boldface letters denote matrices (vectors). $(\cdot)^H$, $(\cdot)^t$, $\mathbf{I}_N$ and $\1$, denote  Hermitian transpose, transpose, $N \times N$ identity matrix and vector of ones, respectively. $\lVert\cdot\rVert$ applied to vectors denotes the Euclidean norm. $\ex{\cdot}$ is the expected value operator. $\circ$ and $\oslash$  denote the Hadamard (pointwise) matrix product and division. $\Re \{\cdot\}$, $\Im\{\cdot\}$, $(.)^*$ and $|\cdot|$ denote the real part, imaginary part, conjugate and absolute value of a complex number, respectively.


\section{System model}
\label{sec:system-model}

Traditional digital modulation schemes transmit information modulating only the amplitude, phase and/or frequency of a sinusoidal carrier. However, Index Modulations (IM) benefit from the fact that the transmitter has several building blocks, being these antennas, polarizations or subcarriers, for example, to map additional bits of information to the block selected to transmit the conventional modulated signal \cite{Basar-2017-IM-NGN}. As illustration, consider  SM with $N_t$ transmit antennas: in addition to the $\log_2(M)$ bits to index each  symbol $s$ in a constellation of $M$ elements,  $\log_2(N_t)$ bits can be used to select which of the $N_t$ antennas is active at a given instant to transmit the symbol. Similarly, PMod, by means of the transmit polarization carries information. In this paper we will focous our attention in the case with $N_t=2$, so that one input bit is used to select the active transmit dimension, although a latter section will explain how to extend the results for $N_t>2$. Hereafter, SM is used instead of IM but keep in mind that results apply to a generic IM, never mind how are interpreted the dimensions (antennas, polarizations, frequencies...).

The system model of a $2\times2$ SM for a given discrete time instant is
\begin{equation}\label{eq:channel_model}
\textbf{y} = \sqrt{\gamma}\textbf{H} \textbf{x} + \textbf{w}
\end{equation}
where $\mathbf{y} \in \mathbb{C}^{2\times 1}$ is the received vector, $\gamma$ the average Signal to Noise Ratio (SNR), $\mathbf{H} \in \mathbb{C}^{2 \times 2}$ the channel matrix,
$\mathbf{x} \in \mathbb{C}^{2\times 1}$ the transmitted signal and $\mathbf{w} \sim \mathcal{CN}(\mathbf{0}, \mathbf{I}_2)$ the Additive White Gaussian (AWGN) noise vector. Since $\mathbf{x}$ has only one component different from zero (component $l$) and its value is $s \in \mathbb{C}$, (\ref{eq:channel_model}) can be also expressed as
\begin{equation}
\textbf{y} = \sqrt{\gamma}\mathbf{h}_l s + \mathbf{w}
\end{equation}
where $\mathbf{h}_l$ denotes the $l$ column of $\mathbf{H}$, $ l \in \{1,2\}$. We assume a unit power constraint, i.e., $\ex{\mathbf{x}^H\mathbf{x}} = \ex{|s^2|}=1$. 

\begin{figure*}[!ht]
	\centering
	\includegraphics[width=0.98\textwidth]{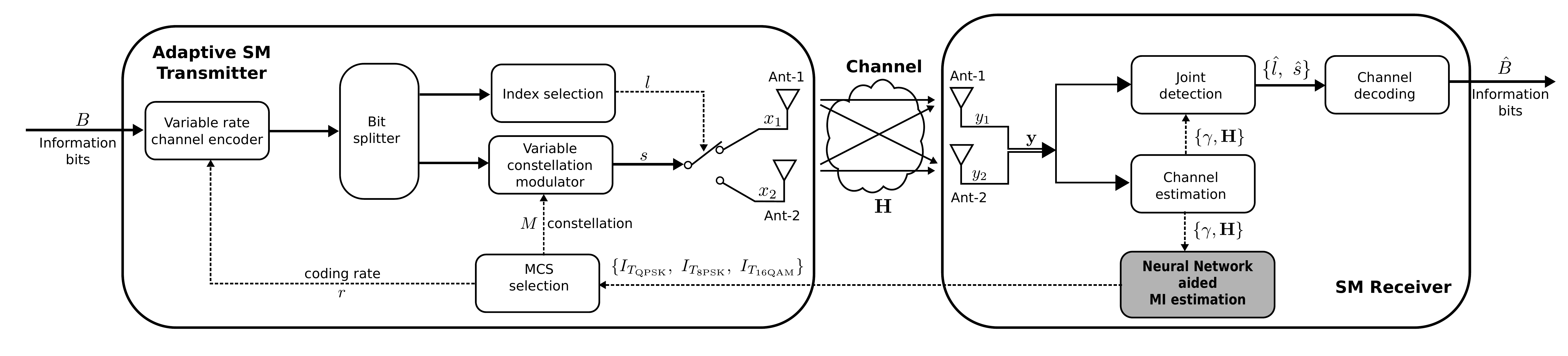}
	\caption{Block diagram of an adaptive Spatial Modulation system with Neural Network aided MI calculation at the receiver.}
	\label{fig:block-diagram}
\end{figure*}

Fig. \ref{fig:block-diagram} shows a block diagram of an adaptive $2\times2$ SM system. The transmitter modifies the coding rate $r$ of the channel encoder and the constellation order $M$ of the modulator according to the different  mutual information (MI) values calculated and reported  by the receiver. In this way, the transmitter adjusts the transmission rate dynamically to the maximum MI that the channel conditions allow at each time instant. Thus, a fine selection of the coding rate is possible due to the accurate calculation of the MI. This approach differs from previous works, \cite{Yang-2015-guidelines-SM} and \cite{Yang-2012-LA-SM}, where the modulation was the only degree of freedom. 

The mapping from the achievable rate (MI) reported by the receiver to the MCSs to be used by the transmitter depends on the strength  of the channel codes and the receiver implementation. In  a practical system a backoff margin should be enforced based on the distance to capacity of the different MCS.  An alternative adaptation method is presented in \cite{ISWCS-2019}. Instead of using the neural network to calculate the MI, the MCS is selected directly. This requires the training of  the network with data from  the MCSs  performance obtained from extensive Monte Carlo  simulations for a vast number of different channel matrices and SNRs.


We should remark again that the model \eqref{eq:channel_model} and the block diagram of Fig. \ref{fig:block-diagram} apply to a generic $2\times2$ IM, with the matrix $\H$ characterizing the channel effects of the specific domain considered, being this polarization, frequency or space. The adaptive SM transmitter requires the knowledge of the MI at each channel realization to perform the link adaptation. In this work we do not deal with the MCS selection at the transmitter, since we focus our attention only on the NN aided MI estimation, denoted by the gray block at the receive side. In order to minimize the overhead of the return link, the receiver estimates the SNR $\gamma$ and the channel matrix $\H$, computes the MI and sends it back to the transmitter. Next section provides the expressions to compute  the MI of a SM system for an arbitrary constellation. A more practical scheme for calculating these MIs, making use of a NN, will be presented later in  Section \ref{sec:MFNN}.

\section{Capacity and Mutual Information}
\label{sec:analytical}

The expression of the Spatial Modulation (SM) capacity conditioned to a given realization of the channel matrix $\mathbf{H}$ is given by 
\begin{equation}
C = \max_{f_X(x)} I(\mathbf{x}; \ \mathbf{y} \ | \mathbf{H}) = \max_{f_S(s), f_L(l)} I(s,l; \ \mathbf{y} \ | \mathbf{H}) \ [bpcu],
\label{eq:general-cap}
\end{equation}
where $I(\x;\y|\H)$ is the Mutual Information (MI) between the two random variables $\x$ and $\y$ conditioned to $\H$, and the maximization is performed for  all  possible distributions of the transmitted signal $\mathbf{x}$ \cite{Cover}. In (\ref{eq:general-cap}), $f_X(x)$, $f_S(s)$ and $f_L(l)$ denote the probability density functions (PDF) of the random variables of the complex transmit signal $\mathbf{x}$, the complex transmit symbol $s$, and the hopping index $l$ which selects the antenna/polarization used to transmit the symbols. The units of the capacity in (\ref{eq:general-cap}) are bits per channel use, \textit{bpcu}. The transmitter is expected to operate with only partial CSIT (it only knows the MI, neither $\H$ nor $\gamma$), so it will select either index $l=\{1,2\}$ with the same probability. The capacity is achieved in \eqref{eq:general-cap} when the  transmit symbols belong to a Gaussian codebook \cite{Narasimhan-2016-letter-GSM}, i.e., when $s \sim \mathcal{CN}(0,1)$.  

The MI in (\ref{eq:general-cap}) can be expressed as a function of the entropies of the involved random variables:
\begin{equation}
I(\mathbf{x}; \ \mathbf{y} \ | \mathbf{H}) = h(\mathbf{y|\mathbf{H}}) - h(\mathbf{y}|\mathbf{x},\mathbf{H}) = h(\mathbf{y|\mathbf{H}}) - h(\mathbf{w}),
\label{eq:general-cap-2}
\end{equation}
where $h(\cdot)$ is used for the differential entropy, and  $h(\mathbf{w})$ is simply written as  
\begin{equation}
h(\mathbf{w}) = \log_2 \det (\pi e \mathbf{I}_2).
\label{eq:entropy-noise}
\end{equation} 
As in \cite{Narasimhan-2016-letter-GSM}, the received vector $\y$ follows a Gaussian distribution of the form 
\begin{equation}
\mathbf{y} \sim \dfrac{1}{2} \sum_{l=1}^{2} \mathcal{CN}\left(\mathbf{0},\ \mathbf{H} \mathbf{K}_l \mathbf{H}^H + \mathbf{I}_2 \right) \triangleq \dfrac{1}{2} \sum_{l=1}^{2} \mathcal{CN}\left(\mathbf{0},\ \mathbf{\Phi}_l \right),
\end{equation}
where  $\mathbf{K}_1 = \begin{pmatrix}
1 & 0\\0 & 0 \end{pmatrix}$ and $\mathbf{K}_2 = \begin{pmatrix}
0 & 0\\0 & 1 \end{pmatrix}$. With this, the entropy of $\y$  in (\ref{eq:general-cap-2}) reads as 
\begin{equation}
h(\mathbf{y}|\mathbf{H}) = - \dfrac{1}{2} \sum_{l=1}^{2} \int_{\mathbf{y}} \mathcal{CN}(\mathbf{0},\mathbf{\Phi}_l) \log_2 \left( \dfrac{1}{2} \sum_{l'=1}^2 \mathcal{CN}(\mathbf{0},\mathbf{\Phi}_{l'}) \right).
\label{eq:entropy-y}
\end{equation}
It is then clear than the computation of the MI $I(\mathbf{x}; \ \mathbf{y} \ | \mathbf{H})$ requires the numerical evaluation of the above integral, which can be too demanding for a receiver updating the estimate of  the link capacity for adaptation purposes. In this paper we will compute the MI by means of Monte Carlo integration as a reference for comparing the approximation performed by the neural network. 

Practical communication links use symbols from a constellation $\mathcal{S}$ with a finite alphabet. Hereafter, we will refer to the corresponding mutual information, or constrained capacity,  simply as total mutual information, $I_T$, since this includes the information carried by both hopping index $l$ and symbol $s$:
\begin{equation}
I_T = \left. I(s,l; \ \mathbf{y} \ | \mathbf{H})\right|_{s \in \mathcal{S}}. 
\end{equation}

The particularization of \eqref{eq:general-cap-2} for a constellation $\mathcal{S}$ with $M$ symbols has been made  in \cite{ICASSP-18-CTTC}, and it is replicated in (\ref{eq:MC-It}) for the sake of completeness. Monte Carlo integration will be also used to compute \eqref{eq:MC-It}, by generating random values of the noise $\w$. 

\begin{figure*}[!ht]
	\normalsize
	\begin{equation}
	I_T = \log_2 (2 M) - \dfrac{1}{2M} \sum_{s \in \mathcal{S}} \sum_{l=1}^{2} \mathbb{E}_{\mathbf{W}} \left \{ \log_2 \left( \sum_{s' \in \mathcal{S}} \sum_{l'=1}^{2} e^{-\gamma \left\| \mathbf{h}_l s - \mathbf{h}_{l'} s' + \dfrac{\mathbf{w}}{\sqrt{\gamma}}\right\|^2 + \left\|\mathbf{w} \right\|^2}    \right) \right \}
	\label{eq:MC-It}
	\end{equation}
\end{figure*}

In an effort to find more convenient expressions to handle in practice, some results have been presented in the literature as approximations to the mutual information \eqref{eq:MC-It}. On the one side, 
Guo et al \cite{Guo-2014-MI-SM-ICC} used the Jensen's inequality and corrected the ensuing bias to get
\begin{equation}
I_{T_{\text{Jensen}}} = -\log_2\left( \dfrac{\sum_{\Delta_x \in \mathcal{D}} e^{-\nicefrac{1}{2} \Delta_x^H \mathbf{H}^H \mathbf{H} \Delta_x} }{(2M)^2} \right).
\label{eq:Guo}
\end{equation}
Here $\mathcal{D}$ is a set with $(2M)^2$ vectors in  $\mathbb{C}^{2\times1}$ of the form 
\begin{equation}
\Delta_x = \sqrt{\gamma}\cdot(\mathbf{h}_l s_k - \mathbf{h}_{l'} s_{k'})
\end{equation}  
for $l,l'=1,2$ and $k,k'=1,2,\dots,M$, where $\mathbf{h}_l$ are the columns of the channel matrix, and $s_k \in \mathcal{S} \subset \mathbb{C}$ the symbols of the constellation.

A different approach resorting to the Taylor Series Expansion was followed in \cite{ICASSP-18-CTTC}, yielding  expression (\ref{eq:Pol}) as an approximation of  $I_T$. The interested reader is referred to \cite{ICASSP-18-CTTC} for the  definitions of each element of the equation. One drawback of both (\ref{eq:Guo}) and (\ref{eq:Pol}) is that the computational complexity of the MI calculation increases with the square of the constellation order $M$ and the number of antennas $N_t$.    

\begin{figure*}[!ht]
	\normalsize
	\begin{equation}
	I_{T_{Taylor}} = \log_2 \left( \dfrac{2M}{\mathfrak{G}(\mathcal{D}_{sl})}\right) + \mathfrak{A} \left( \dfrac{\log_2 \left( \mathfrak{G}_{sl} \left( {D_{s,l,s',l'}}^{D_{s,l,s',l'}} \right) \right)}{\mathfrak{A}_{sl}\left(D_{s,l,s',l'} \right)} + \dfrac{\gamma}{\log(2) \mathcal{D}_{sl}^2} \sum_{m=1}^{2} \left( \mathcal{D}_{m,sl,\mathfrak{R}}^2 + \mathcal{D}_{m,sl,\mathfrak{I}}^2  \right)\right)
	\label{eq:Pol}
	\end{equation}
\end{figure*}

This paper develops a more efficient and accurate scheme to compute the MI of a $2\times 2$ SM system, which avoids the quadratic complexity increment with the constellation cardinality. This is especially relevant for practical use, given the need to estimate on the fly the channel capacity for link adaptation purposes. In an adaptive coding and modulation (ACM) system, the receiver must feed back to the transmitter a metric related to the achievable rate, so that the transmitter can select the appropriate MCS\footnote{The receiver itself can also make this choice and report back the corresponding index to the transmitter.}. This estimation process needs to be both simple and accurate, since errors will lead to wrong choices of MCSs, and lower rates than supported or decoding mistakes will occur. The proposed scheme to estimate the achievable rate is  based on a simple NN with only one hidden layer, and which provides different outputs, one per constellation in case several are available. The computational burden is much lower than that of any other previously known alternatives, so the MI can be updated more often and faster variations of the channel conditions can be tracked as a benefit. 




\section{Neural Network-based Mutual Information Estimation}
\label{sec:MFNN}

\begin{figure}[!ht]
	\centering
	\includegraphics[width=0.98\columnwidth]{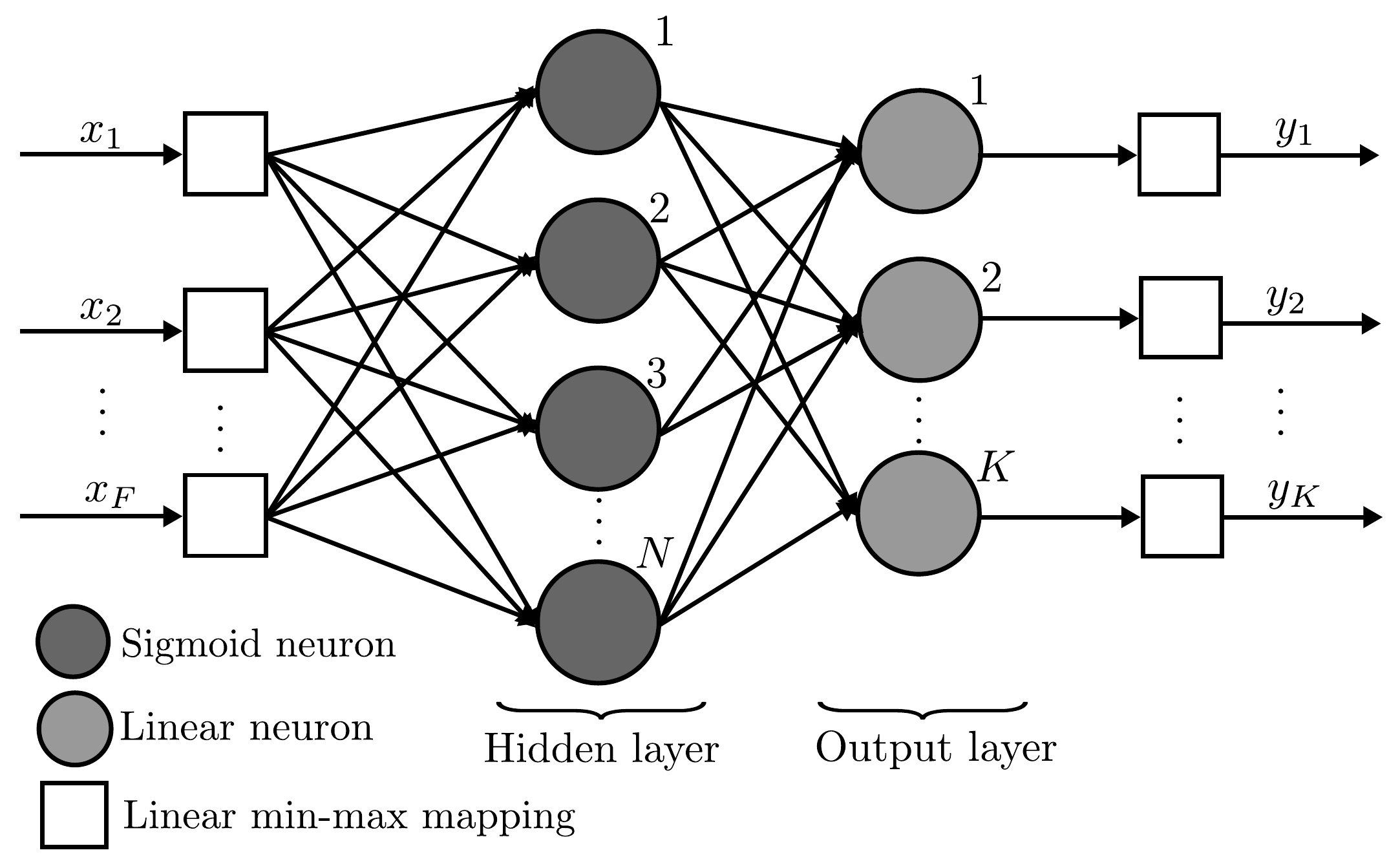}
	\caption{Diagram of the neural network.}
	\label{fig:MFNN}
\end{figure}

The evaluation of the Mutual Information (MI)  \eqref{eq:MC-It} can be interpreted as a non-linear mapping from the channel matrix $\H$ and the SNR  $\gamma$ to the MI. Multilayer Feedforward Neural Networks (MFNNs), well-known for their fitting  capabilities of non-linear functions   \cite{Funahashi-1989}-\cite{Hornik-1991}, will be used to estimate $I_T$ in \eqref{eq:MC-It}. In particular, the MFNN to be employed, a one hidden layer network, is detailed in Fig. \ref{fig:MFNN}. The input features,  $\x=[x_1,x_2,\dots,x_F]^t$, need to be extracted by means of a function $f(\cdot)$ from the   channel matrix $\H$ and the SNR $\gamma$, as $\x=f(\gamma,\H)$. The input variable selection is highly relevant for the performance of the learning process of the network. Later we will detail how the \textit{feature  extraction}  is applied  based on our \textit{domain knowledge}, that is, our knowledge of the particular problem we are addressing. Alternatively, a more complex deep neural network with several hidden layers could be used, so that the relevant features for this problem are learned  by the intermediate layers, and all the channel matrix coefficients are used directly as inputs without any processing. However, as detailed later, we found this solution underperforming the one hidden layer with carefully selected input features. 

In the following, the variables in blue will denote the internal parameters of the neural network, and $\a_i,i=0,1,2$ will be the intermediate internal variables at the $i$-th layer. Each of the $F$ neural network inputs goes through a linear preprocessing block to adjust the neurons input to the range $[-1,+1]$. This initial scaling is expressed as 
\begin{align}\label{eq:a0}
\a_0 = \textcolor{blue}{\mathbf{g}_0} \circ (\mathbf{x}-\textcolor{blue}{\mathbf{x}_0}) - \1 \ \in \mathbb{R}^{F\times1}
\end{align}
with the gain $\mathbf{g}_0 \in \mathbb{R}^{F\times1}$ and the offset $\mathbf{x}_0 \in \mathbb{R}^{F\times1}$. 

The hidden layer is made of $N$ neurons, also named processing units, each applying a weighted linear combination of its inputs, a bias and a non-linear function, also known as activation function:
\begin{align}
\a_1 = g (\textcolor{blue}{\mathbf{W}_1} \cdot \a_0 + \textcolor{blue}{\mathbf{b}_1}) \in \mathbb{R}^{N\times1}.
\end{align}
The matrix $\mathbf{W}_1 \in \mathbb{R}^{N\times F}$ and the vector $\mathbf{b}_1 \in \mathbb{R}^{N\times1}$ collect the weights and the offsets. As activation function we will use the hyperbolic tangent:
\begin{equation}
\nonumber
g(x) = \dfrac{2}{1+e^{-2x}}-1.
\end{equation}
The output layer of $K$ neurons applies a linear processing of the form  
\begin{align}
\a_2 = \textcolor{blue}{\mathbf{W}_2} \cdot \a_1 + \textcolor{blue}{\b_2} \ \in \mathbb{R}^{K\times1}
\end{align}
for matrix $\mathbf{W}_2 \in \mathbb{R}^{K\times N}$ and  vector $\b_2 \in \mathbb{R}^{K\times1}$. Finally, there is a last stage to accommodate the range of the network outcome:
\begin{align}\label{eq:y}
\y = (\a_2 +\1) \oslash \textcolor{blue}{\g_3} + \textcolor{blue}{\y_0} \ \in \mathbb{R}^{K\times1}
\end{align}
with the gain $\g_3  \in \mathbb{R}^{K\times1}$ and the offset $\y_0  \in \mathbb{R}^{K\times1}$.
The output vector is expressed as $\y=[y_1,y_2,\dots,y_K]^t$.

The different parameters of the network --weights and biases in \eqref{eq:a0}-\eqref{eq:y}--  will be obtained under supervised learning. The Levenberg-Marquardt (LM) backprogation algorithm \cite{Hagan-1994} will be used to minimize the Mean Squared Error (MSE) on the test set:
\begin{equation}
\text{MSE} =  \frac{1}{L} \sum_{\ell=1}^L \lVert \y(\ell) - \t(\ell) \rVert^2, 
\end{equation}
with $\y(\ell)$ and $\t(\ell)$ the network output and the true MI values, respectively, for the training tuples  $(\x(\ell),\t(\ell))$, $\ell=1,\ldots,L$. The true MI  values are obtained by numerical evaluation of \eqref{eq:MC-It} for $K$ different symbol constellations. The training of the network is performed off-line, so that receiver terminals only need to evaluate \eqref{eq:a0}-\eqref{eq:y} for adaptation purposes. The computational complexity of the NN will be evaluated later in comparison to that of previous methods presented in the literature \cite{ICASSP-18-CTTC}-\cite{Guo-2014-MI-SM-ICC}.

Hereafter, we focus on the MI calculation of a $2\times2$ SM system and next section will provide simulation results for this particular case. However, the underlying philosophy applies to SM with a higher number of antennas, not only $2$. Section \ref{sec:extension} will show how to extend the results to obtain the MI of SM systems with a larger number of antennas. 
Remarkably, the neural network entails always a degree of complexity about two orders of magnitude lower compared to that of the  analytical approximations, even for SM systems with high number of antennas/dimensions.  


\subsection{Input Variable Selection}

The output of the network in Fig. \ref{fig:MFNN} is the MI estimation for $K$ different constellations, as an approximation of the true MI function \eqref{eq:MC-It}. This depends on the channel matrix $\H$ and the SNR $\gamma$; in the following we will see how to pre-process these values for a better network performance. 

For a Maximum Likelihood receiver, the pairwise error probability (PEP) between $(s,l)$  and $(\hat{s},\hat{l})$ is given by \cite{Yang-2012-LA-SM},\cite{Zafari-DP-SM}
\begin{equation}
P_e(s,\hat{s},l,\hat{l}) = Q \left( \sqrt{\dfrac{\gamma}{2}}  \lVert \h_l \cdot s - \h_{\hat{l}} \cdot \hat{s} \rVert \right)
\end{equation}
for the AWGN case. This PEP depends on the distance among supersymbols $\h_l \cdot s$. The set $\mathcal{X}$ of  $2M$ different supersymbols for a given channel matrix $\H$ and the transmitted constellation $\mathcal{S} = \{s_k, \ k=1,2,\dots,M\}$ is 
\begin{equation}
\mathcal{X} = \{\h_l s_k,\ l=1,2,\ k=1,2,\dots,M\}.
\end{equation}
The MI will be also affected by all the involved distances, as inferred from \eqref{eq:MC-It}. For convenience, we put together the squared distances among all the pairs of supersymbols under the matrix $\D$ of size $2M\times 2M$, with the respective entries given by 
\begin{equation}
\D[(l-1)M+k,(l'-1)M+k'] = \lVert \h_l s_k - \h_{l'} s_{k'}\rVert^2
\end{equation}
where $l,l'=\{1,2\}$ and $k,k'=\{1,2,\dots,M\}$. Matrix $\D$ can be also expressed as 
\begin{equation}
\D = 
\begin{pmatrix}
\lVert \h_1 \rVert^2 \D_S   & \D_L \\ 
\D_L^t &  \lVert \h_2 \rVert^2 \D_S
\end{pmatrix}.
\end{equation}
The  $M\times M$ matrix $\D_S$ on the diagonal is a function of the symbols in the constellation  $\mathcal{S}$:
\begin{equation}
\D_S[k,k'] = |s_k - s_{k'}|^2, 
\end{equation}
whereas the $M\times M$ matrix $\D_L$  contains all the distances between supersymbols of different antennas/polarizations:
\begin{equation}
\begin{aligned}
\D_L[m,n] & = \lVert\h_1 s_m - \h_2 s_n\rVert^2 \\
& =  \lVert\h_1 s_m\rVert^2 + \lVert\h_2 s_n\rVert^2 - 2 \Re\{s_m^*s_n \h_1^H \h_2\}.
\end{aligned}
\label{eq:DL-first}
\end{equation}

Note that the matrix $\D_L$ can be expressed as the sum of four rank-$1$ matrices: 
\begin{multline}
\D_L = \lVert\h_1\rVert^2 \left(\begin{array}{c} |s_1|^2 \\ \vdots \\  |s_M|^2
\end{array} \right) \1^t + \lVert\h_2\rVert^2 \; \1  
\left(\begin{array}{ccc} |s_1|^2 & \hdots &    |s_M|^2
\end{array} \right)  \\ 
- \h_1^H \h_2 \left(\begin{array}{c} s_1^* \\ \vdots \\ s_M^*
\end{array} \right) \left(\begin{array}{ccc} s_1 & \hdots & s_M
\end{array} \right) \\
- \h_1^t \h_2^* \left(\begin{array}{c} s_1 \\ \vdots \\ s_M
\end{array} \right) \left(\begin{array}{ccc} s_1^* & \hdots & s_M^*
\end{array} \right).
\label{eq:Dl}
\end{multline}
With this, we have that $\text{rank}\{\D_L\} \leq 4$. Even further, if the constellation of symbols $\{s_n\}$ is known, then only four  real values are required to describe the dependence of $\D_L$ and $\D$ with the channel matrix $\H$, namely, $\lVert\h_1\rVert^2,\lVert\h_2\rVert^2$ and the real and imaginary parts of the scalar product $\h_1^H \h_2$. Alternatively, the scalar product can be expressed as \cite{Scharnhorst-2001} 
\begin{equation}
\h_1^H \h_2 = \lVert \h_1 \rVert \cdot \lVert \h_2 \rVert
\cdot \cos \Theta_H \cdot e^{i\varphi} 
\end{equation}
where $\Theta_H \in [0,\pi/2]$ and $\varphi \in [-\pi,\pi]$ denote, respectively, the Hermitian angle and the Kasner's pseudo-angle between two complex vectors. Thus, the four values $(\lVert \h_1 \rVert, 
\lVert \h_2 \rVert, \Theta_H, \varphi)$ serve to characterize the matrix $\D$. 

\begin{figure*}[!ht]
	\centering
	\subfloat[Orthogonal columns, SNR$=10$ dB.]{
		\includegraphics[width=0.3\textwidth]{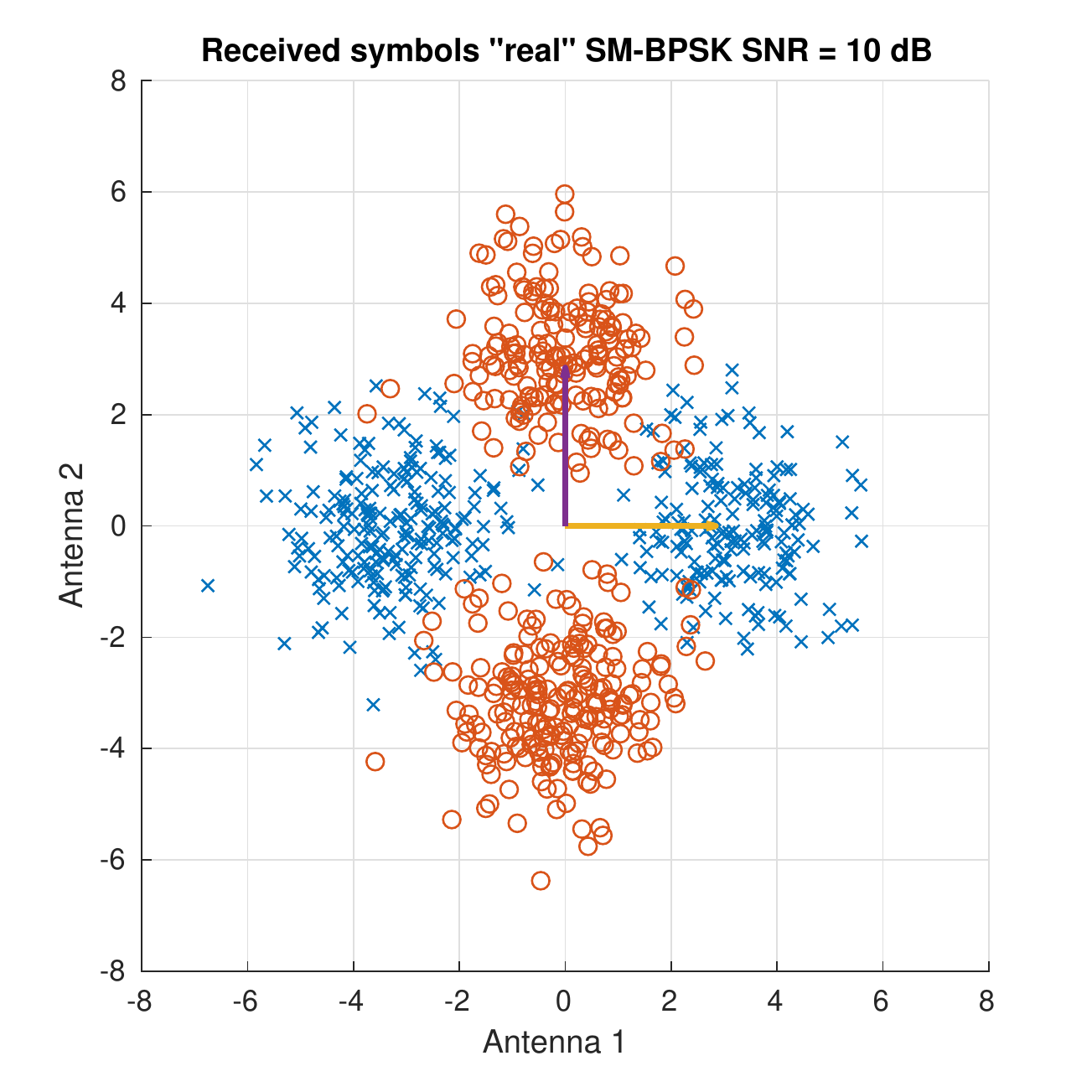}}
		\label{sfig:bpsk_ort_10}
	\hfil
	\subfloat[Orthogonal columns, SNR$=15$ dB.]{
		\includegraphics[width=0.3\textwidth]{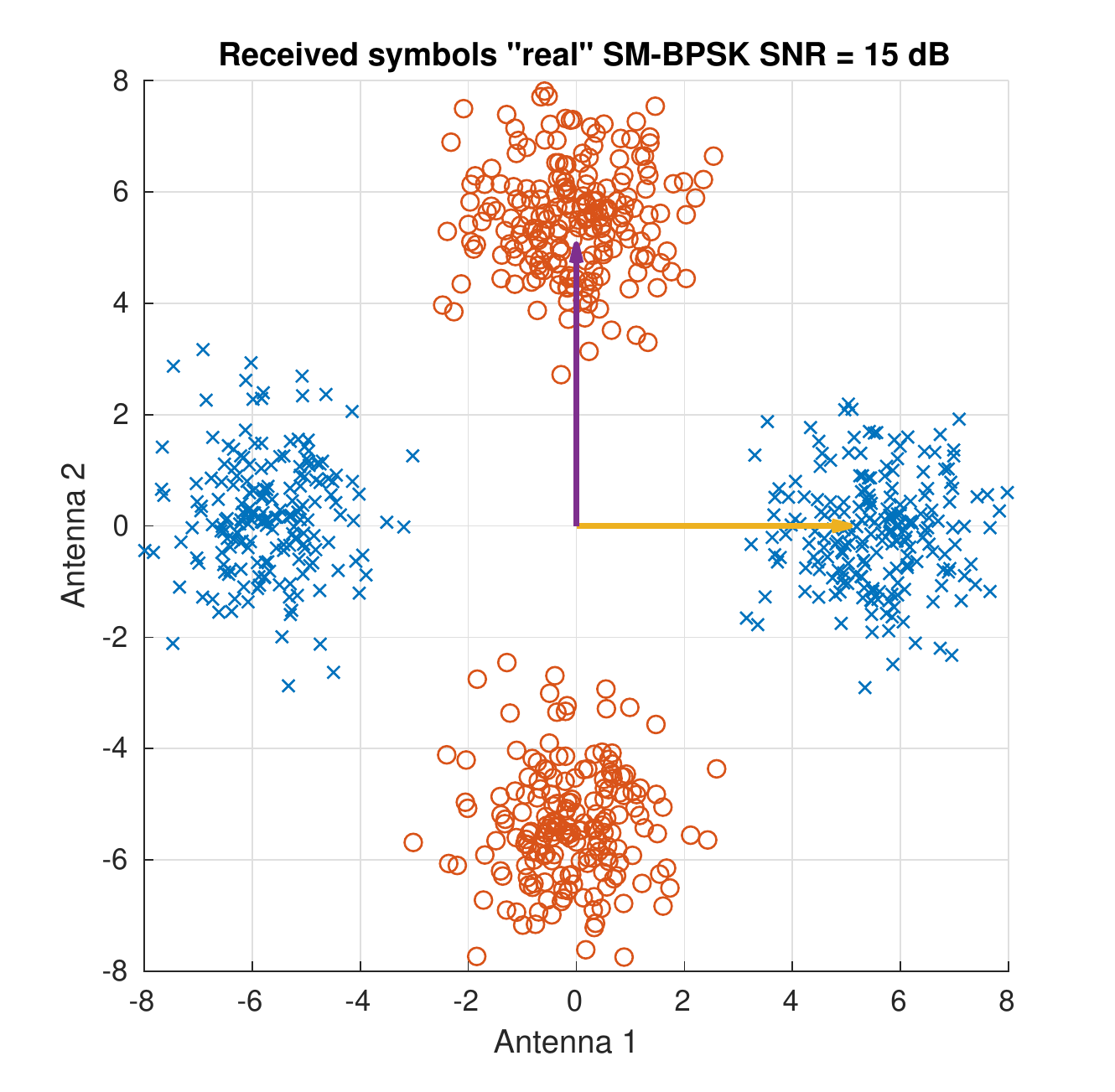}} \\
		\label{sfig:bpsk_ort_15}
	\subfloat[Non-orthogonal columns, SNR$=10$ dB.]{
		\includegraphics[width=0.3\textwidth]{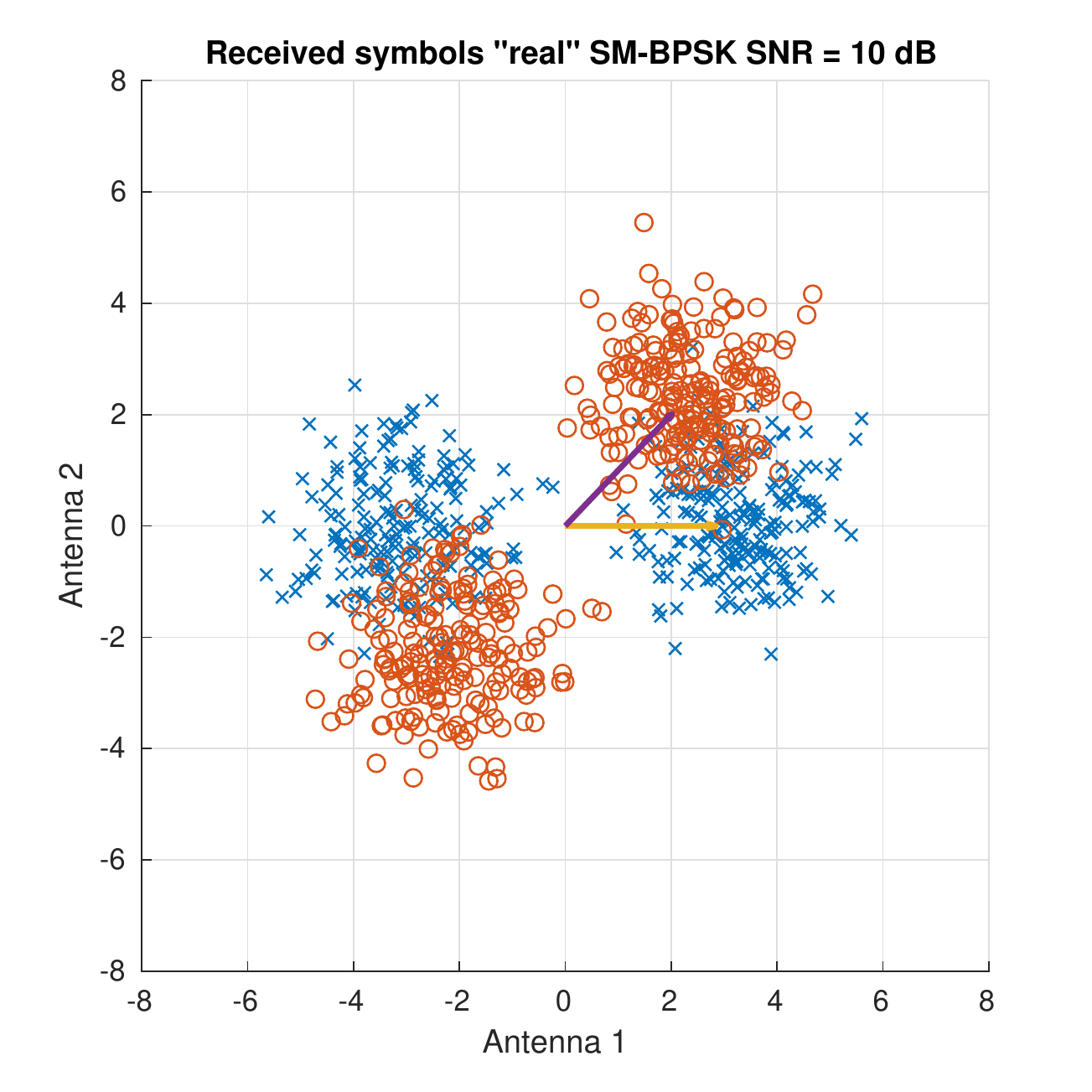}
		\label{sfig:bpsk_nort_10}}
    \hfil
	\subfloat[Non-orthogonal columns, SNR$=15$ dB.]{
		\includegraphics[width=0.3\textwidth]{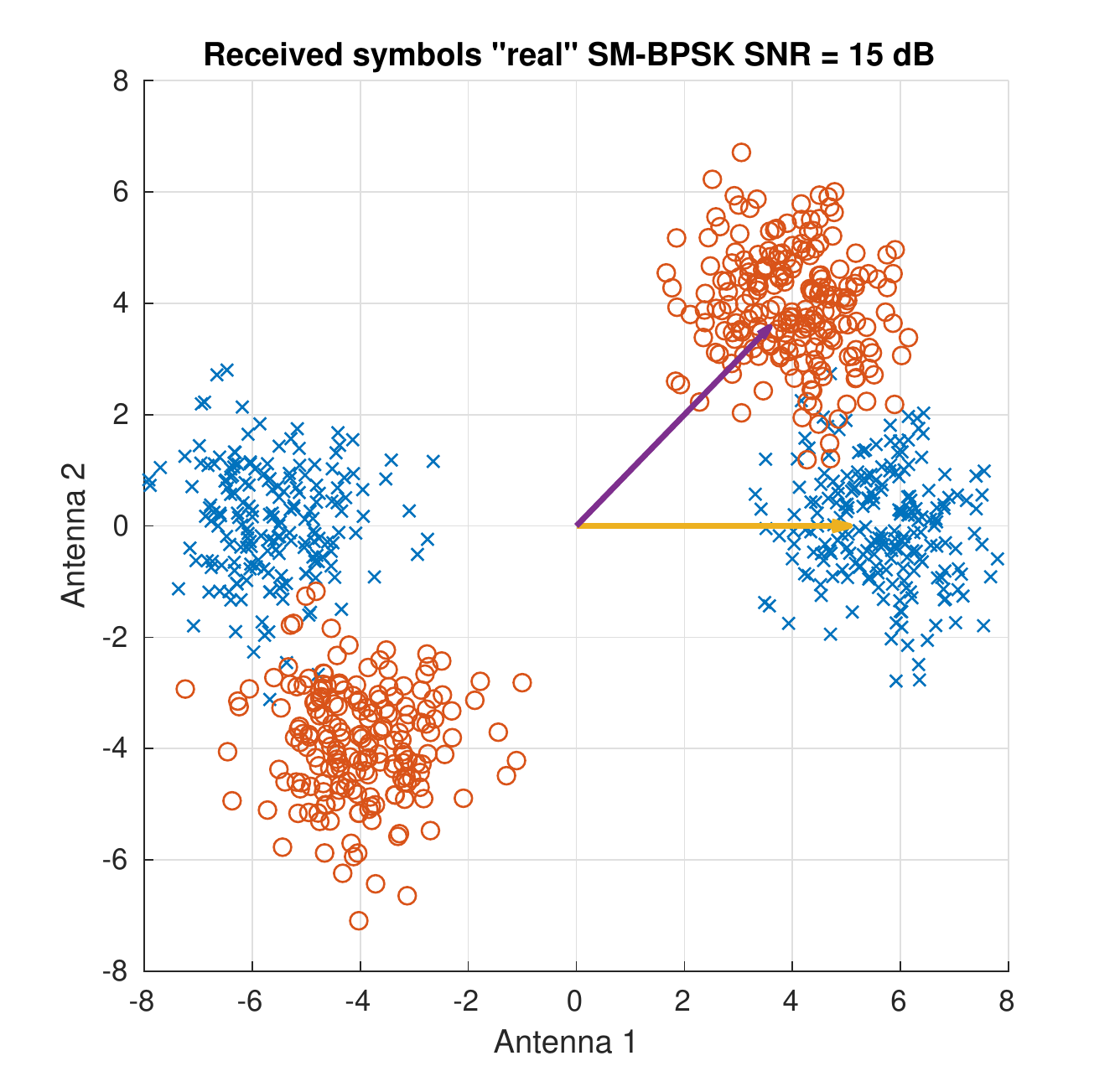}
		\label{sfig:bpsk_nort_15}}
	
	\caption{Received constellation for $2\times2$ SM-BPSK system where transmitted symbols, channel matrix and noise are real-valued.}
	\label{fig:bpsk}
\end{figure*}

For illustration purposes, Fig. \ref{fig:bpsk} shows the received symbols for a real BPSK case. Two different SNR values and two different channel matrices are employed to display the clouds of received symbols. Both the SNR and the angle between the column vectors of $\H$ determine the distance among the different color clouds. 

\begin{figure}[!h]
	\centering
	\includegraphics[width=0.99\columnwidth]{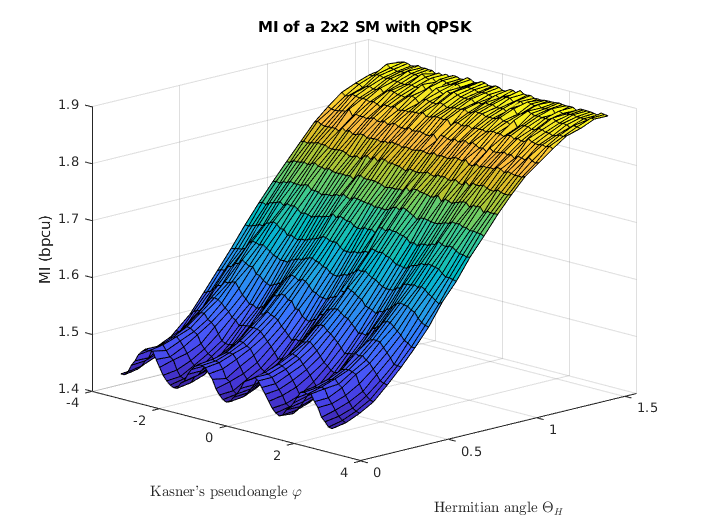}
	\caption{MI of a $2\times2$ SM link with QPSK constellation as a function of the two angles for unit-valued columns norms and $3$ dBs of SNR.}
	\label{fig:mi-3d}
\end{figure}

The impact of the two angles $\Theta_H$ and $\varphi$ in the final MI can be grasped with the aid of Fig. \ref{fig:mi-3d}, which shows a 3D 
 representation of the MI as function of both angles. Monte Carlo simulations were run for a QPSK constellation, with  both columns having the same norm, $\lVert \h_1 \lVert = \lVert \h_2\lVert$ = 1, and $\gamma=2$. With this, the structure of the  channel matrix $\H$ is the following:
\begin{equation}
\H = \begin{pmatrix}
1 & \cos \Theta_H e^{i \varphi}\\
0 &  \sin \Theta_H
\end{pmatrix}.
\end{equation}

It can be seen that the MI has a strong dependance with the Hermitian angle. If $\Theta_H=\pi/2$ the two columns are orthogonal and the MI is maximum, whereas for  $\Theta_H=0$ the columns are considered parallel,  and the MI is reduced. Moreover, the Kasner's pseudoangle $\varphi$ only affects the MI significantly when $\Theta_H$ is close to zero,  creating a small ripple, due to the radial symmetry of the constellation. However, for $\Theta_H > \pi/3$,  the MI is barely affected by the Kasner's pseudoangle.  Instead, if $\Theta_H=0,$ the phase $\varphi$ determines to which extent the receiver can tell which antenna transmitted the observed symbol. 

After extensive training cases, we have observed that performance can be enhanced if, as part of the input parameters, the distances among the supersymbols are also included. Four distances, $\{d_i\},i=1,\ldots,4$, are used; this is the number of different entries of matrix  
 $\D_L$ in  (\ref{eq:DL-first}) when a QPSK constellation is employed. It turns out that these four quantities suffice for the neural network to compute a good estimate of the MI for other constellations too, such as 8PSK and 16QAM, even though the number of different entries of the matrix is higher. 
 

\subsection{Neural Network Operation}

We will use a unique  MFNN to estimate the mutual information $I_T$ for different symbol constellations. $K=3$ outputs will provide   the estimate of the mutual information for QPSK, 8PSK and 16QAM constellations. Following the above considerations, the input features
$\x = f(\gamma, \H)$ will correspond to different options to characterize the distance matrix $\D$. Table \ref{tab:features} depicts the different feature sets that will be used for testing the performance of the network. 
Essentially, four inputs is the lowest number of inputs to test following the previous discussion\footnote{Note that the number of real values to characterize $\H$ and $\gamma$ is nine.}. Values are sorted in ascending order given the invariance of the capacity to the labeling of the dimensions and symbols. 
As to the number of neurons, $N=10$ and $N=20$ will be used in the simulations.

\begin{table*}[!ht]
	\centering
	\begin{tabular}{c|c|l}
		\textbf{Option} & \textbf{Input features $F$} & \textbf{Description of the features} \\ \hline
		i & $4$ & $\x = \left[  \ \text{sort}\left([\gamma \lVert \h_1 \rVert^2, \ \gamma \lVert \h_2 \rVert^2]\right), \ \Re \left \{ \dfrac{\h_1^H \h_2}{\lVert \h_1 \rVert \cdot \lVert \h_2 \rVert} \right \}, \ \Im \left \{ \dfrac{\h_1^H \h_2}{\lVert \h_1 \rVert \cdot \lVert \h_2 \rVert} \right \} \right]^t$ \\ \hline
		ii & $4$ & $\x = \left[  \ \text{sort}\left([\gamma \lVert \h_1 \rVert^2, \ \gamma \lVert \h_2 \rVert^2]\right), \ \Theta_H, \ \varphi \right]^t$ \\ \hline
		iii & $6$ & $\x = \left[  \ \text{sort}\left([\gamma \lVert \h_1 \rVert^2, \ \gamma \lVert \h_2 \rVert^2]\right),\  \text{sort}([\gamma d_1, \ \gamma d_2, \ \gamma d_3, \ \gamma d_4])  \right]^t$ \\ \hline
		iv & $8$ & $\x = \left[  \ \text{sort}\left([\gamma \lVert \h_1 \rVert^2, \ \gamma \lVert \h_2 \rVert^2]\right),\  \text{sort}([\gamma d_1, \ \gamma d_2, \ \gamma d_3, \ \gamma d_4]),\  \Re \left \{ \dfrac{\h_1^H \h_2}{\lVert \h_1 \rVert \cdot \lVert \h_2 \rVert} \right \}, \ \Im \left \{ \dfrac{\h_1^H \h_2}{\lVert \h_1 \rVert \cdot \lVert \h_2 \rVert} \right \} \right]^t$ \\ \hline
		v & $8$ & $\x = \left[  \ \text{sort}\left([\gamma \lVert \h_1 \rVert^2, \ \gamma \lVert \h_2 \rVert^2]\right),\  \text{sort}([\gamma d_1, \ \gamma d_2, \ \gamma d_3, \ \gamma d_4]), \ \Theta_H, \ \varphi  \right]^t$ \\ \hline
		
	\end{tabular}
	\caption{Different alternatives for selecting the NN input features.}
	\label{tab:features}
\end{table*}

Training of the network will be based on extensive amount of data, by generating a large number or random channels for different values of SNR. The reference true capacity values for the different constellations will be obtained after computing \eqref{eq:MC-It} by the Monte Carlo method. 




\section{Simulation results}
\label{sec:results}

For performance evaluation, a  dataset  of $50,000$ realizations of the channel matrix $\mathbf{H}$ is used, randomly generated  following a unit-variance Rayleigh distribution, i.e., $h_{ij} \sim \mathcal{CN}(0,1)$. Each channel matrix is associated with a different SNR whose value in decibels is drawn from a uniform random variable between $-20$ and $20$ dB. The true MI with QPSK, 8PSK and 16QAM constellations of each realization of ($\gamma,\mathbf{H}$) is calculated with a Monte Carlo simulation using (\ref{eq:MC-It}) with $5,000$ realizations of the complex Gaussian noise $\mathbf{w}$. 
We limit ourselves to these low order constellations, which are more likely to be used with SM. However, other constellations, like 64QAM, could be easily added to the system at the expense of increasing the time required for obtaining the dataset with Monte Carlo simulations -note the two summations over all the constellation symbols in (\ref{eq:MC-It}).

The dataset is divided into two independent parts. $7,500$ samples ($15 \%$) are reserved  for the final test of the performance of the MFNN and the analytical approximations (equations (\ref{eq:Guo} and (\ref{eq:Pol})). The remaining $35,000$ samples ($70 \%$) and $7,500$ samples ($15 \%$) are employed for training and validation of the neural network, respectively.  

\begin{figure}[!ht]
	\centering
	\includegraphics[width=0.98\columnwidth]{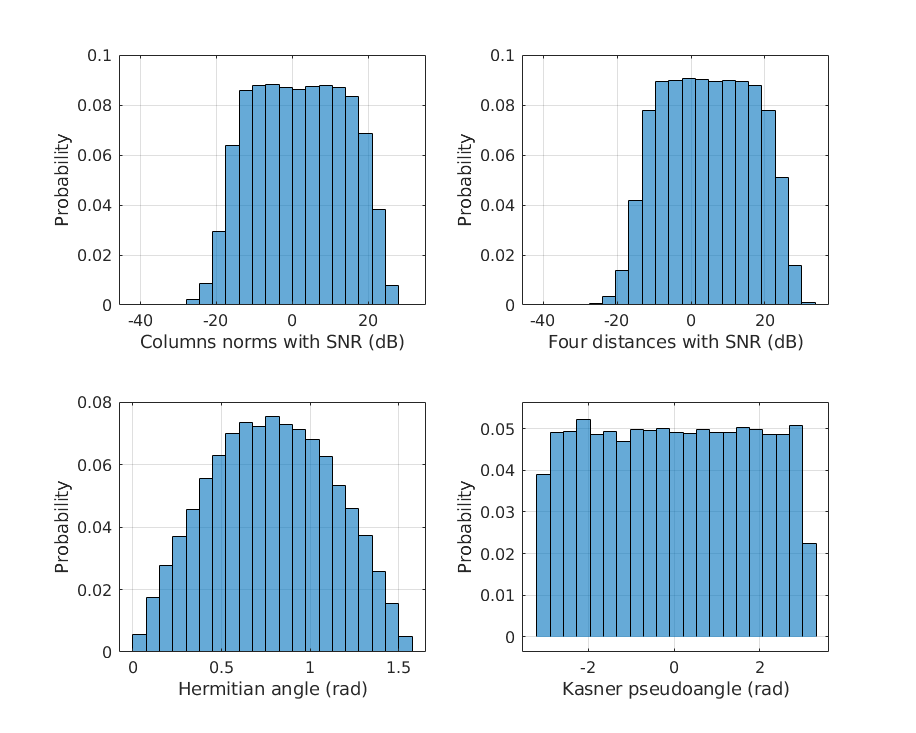}
	\caption{Histograms with the distribution of the features (norms, distances and angles) in the dataset of Rayleigh distributed channel matrices and uniformly distributed SNR.}
	\label{fig:hists}
\end{figure}

Firstly, the impact of the selection of the input features in the performance of the MFNN is evaluated. For this, several NNs are trained using in each case one of the sets of inputs detailed in Table \ref{tab:features}. Fig. \ref{fig:hists} shows some histograms with the statistical distribution of the features obtained from the dataset. The distances and the norms include the SNR term and are shown in dB, whilst the unit of the angles is the radian.

Then, the global MSE obtained with the trained NNs when calculating the three MI values is obtained by using the entries of the dataset reserved for testing. Table \ref{tab:mse-NNs} collects the values of MSE for five different selections of the input features and for both numbers of neurons ($N=10$ and $N=20$). It shows the best MSE in the testing dataset after $10$ trainings with different NN parameters initialization. If the NNs are fed directly with the real and imaginary parts of the channel matrix coefficients, the MSE is very high, in the order of $10^ {-1}$. Nevertheless, at least two orders of magnitude improvement is achieved when the NNs are fed with the input features detailed in Table \ref{tab:features}.

\begin{table*}[!ht]
	\centering
	\begin{tabular}{|l|c|c|c|}
		\hline
		\textbf{Input features option} & \textbf{ Number of features $F$} & \textbf{Global MSE (10 neurons)} & \textbf{Global MSE (20 neurons)} \\ \hline
		i) Columns norm and projection & $4$ & $1.78 \cdot 10^{-3}$  & $6.98 \cdot 10^{-4}$\\ \hline
		ii) Columns norm and angles & $4$ & $4.29 \cdot 10^{-4}$  & $3.36 \cdot 10^{-4}$\\ \hline
		iii) Columns norm and distances & $6$ & $1.79 \cdot 10^{-4}$  & $5.21 \cdot 10^{-5}$\\ \hline
		iv) Columns norm, distances and projection & $8$ & $1.33 \cdot 10^{-4}$  & $4.96 \cdot 10^{-5}$\\ \hline
		v) Columns norm, distances and angles & $8$ &  $1.00 \cdot 10^{-4}$  &  $2.97 \cdot 10^{-5}$\\ \hline
	\end{tabular}
	\caption{Comparison of the global MSE obtained with the neural network for different input features.}
	\label{tab:mse-NNs}
\end{table*}

In Table \ref{tab:mse-NNs}, it can be observed how the Hermitian angle $\Theta_H$ and the Kasner's pseudoangle $\varphi$, options (ii) and (v), improve the NN estimation with respect to the use of the  real and imaginary parts of the projection, options (i) and (iv). Furthermore, the addition of  the four distances to the set of inputs, cases (iii), (iv) and (v), serves to reduce the  MSE as compared to cases (i) and (ii). Finally, the  MSE reaches a minimum value of about $3\cdot10^{-5}$ when the four distances and the two column norms are combined with the two angles for the $20$ neurons MFNN.

Secondly, the two NNs with $10$ and $20$ neurons and the input features selection (v), are compared with the analytical approximations from the literature, (\ref{eq:Guo}) and (\ref{eq:Pol}), in Table \ref{tab:comparison}.  Both Taylor and Jensen based approximations have a similar MSE, around $10^{-2}$, which is outperformed  by all the NN reported in Table \ref{tab:mse-NNs}. Moreover, when we compare the analytical approximations with the best NNs of the table, the improvement in the MSE is about $100$ and $600$ times with a NN of $10$ and $20$ neurons, respectively. 


As noted previously, the calculation of the MI could be addressed with a deep neural network, a MFNN with several  hidden layers, using as inputs the channel matrix coefficients (scaled by the SNR) directly. With this approach, the network extracts the relevant features at the intermediate layers, so that the last layer computes the MI.   We have tested this approximation for a number of layers ranging from one to ten, and a number of neurons per layer between $20$ and $50$.  It was found that  a deep network with at least three layers of $20$ neurons can perform better than the Taylor and Jenson approximations, yielding an MSE in the order of $10^{-3}$. However, the deep networks do not overcome the peformance of the one-hidden layer MFNN which the input features of row (v) of Table \ref{tab:features}. In addition, the training of these deep networks is much more time consuming, and the learning algorithm has more difficulties to converge to those  parameter values providing  a good performance.

As shown in Table \ref{tab:comparison}, the analytical approximation suffers a maximum error of $0.741$, in the 16-QAM constellation,  which is reduced to $0.105$ in the case of the MFNN with $20$ neurons. The improvement is even more noticeable with the $3\sigma$ value: a little bit more than $0.300$ for the analytical approximations, and $20$ times smaller with the neural network.

\begin{figure*}[!ht]
	\centering
	\subfloat[Taylor approximation (\ref{eq:Pol}).]{
		\includegraphics[width=0.32\textwidth]{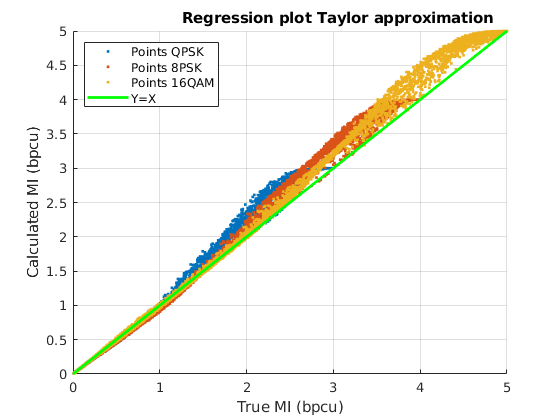}
		\label{sfig:scatt-1st}}
	\subfloat[Jensen based approximation (\ref{eq:Guo}.)]{
		\includegraphics[width=0.32\textwidth]{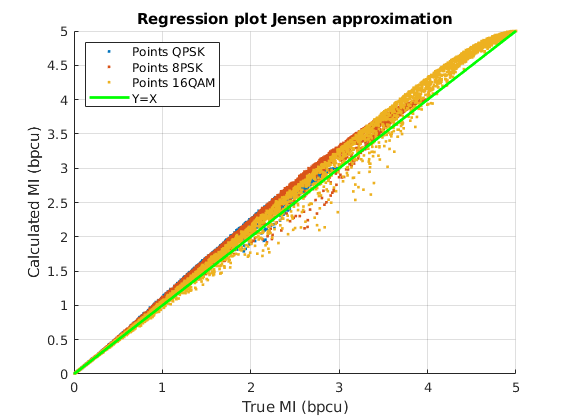}
		\label{sfig:scatt-2nd}} 
	\subfloat[Neural network option (v), 10 neurons.]{
		\includegraphics[width=0.32\textwidth]{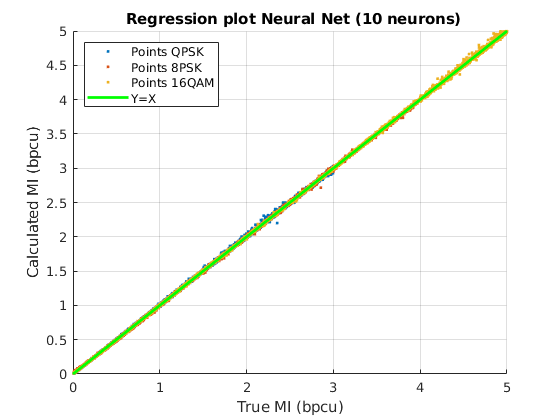}
		\label{sfig:scatt-an}}
	\hfil
	\caption{Comparison of the scatter plots (true MI vs calculated MI) of the analytical approximation from the literature and the MFNNs with $10$ neurons for the three constellations (QPSK, 8PSK and 16QAM).}
	\label{fig:scatterplots}
\end{figure*}

\begin{table*}[]
	\centering
	\begin{tabular}{|l|c||c|c||c|c||c|c|}
		\hline
		\multirow{2}{*}{} & \multirow{2}{*}{\textbf{Global MSE}} & \multicolumn{2}{c||}{\textbf{QPSK}} & \multicolumn{2}{c||}{\textbf{8PSK}} & \multicolumn{2}{c|}{\textbf{16QAM}} \\ \cline{3-8} 
		&  & $\mathbf{3\sigma}$ & \textbf{Max. error} & $\mathbf{3\sigma}$ & \textbf{Max. error} &  $\mathbf{3\sigma}$ & \textbf{Max. error} \\ \hline
		\textbf{Taylor approximation} (\ref{eq:Pol}) & $1.87 \cdot 10^{-2}$  & $0.330$  & $0.523$   & $0.370$ & $0.492$ & $0.392$  & $0.558$  \\ \hline
		\textbf{Jensen based approximation} (\ref{eq:Guo}) & $1.21 \cdot 10^{-2}$   & $0.229$   & $0.300$   & $0.291$  & $0.498$ & $0.300$  & $0.741$  \\ \hline
		\textbf{MFNN option (v) with 10 neurons} & $1.00 \cdot 10^{-4}$  & $0.020$ & $0.153$  & $0.026$  & $0.140$   & $0.034$  & $0.120$  \\ \hline
		\textbf{MFNN option (v) with 20 neurons} &  $2.97 \cdot 10^{-5}$  & $0.016$ &  $0.067$  &  $0.015$  &  $0.046$  &  $0.018$  &   $0.105$  \\ \hline
	\end{tabular}
	\caption{Comparison of the performance of the MFNN with the analytical approximations of the literature for calculating the MI of a $2\times2$ SM.}
	\label{tab:comparison}
\end{table*}
Fig. \ref{fig:scatterplots} shows a graphical view of the estimated MI values: the scatter plot of the values of the true MI (X axis) are shown together with the values of the MI computed with each method (Y axis) for the three constellations, QPSK, 8PSK and 16QAM. The green line Y=X sets the perfect match of the MI. It can be seen that the analytical approximations provide better results for lower values of MI, while for MIs close to their maximum ($3$, $4$ or $5$, depending on the constellation), they have a noticeable positive bias. Remarkably, both NNs with $10$ and $20$ neurons in the hidden layer, match the true value of the MI almost perfectly, clearly outperforming the analytical approximations.

The accuracy achieved by the MFNN has a direct impact on the implementation of adaptive IM links. The quality of the tracking of the MI allows to use smaller back-off margins for the selection of the MCS; large errors make it necessary to use highly conservative margins in the selection of the MCS to guarantee a prescribed error decoding metric, thus reducing the transmission rate. 

\begin{figure}[!ht]
	\centering
	\includegraphics[width=0.85\columnwidth]{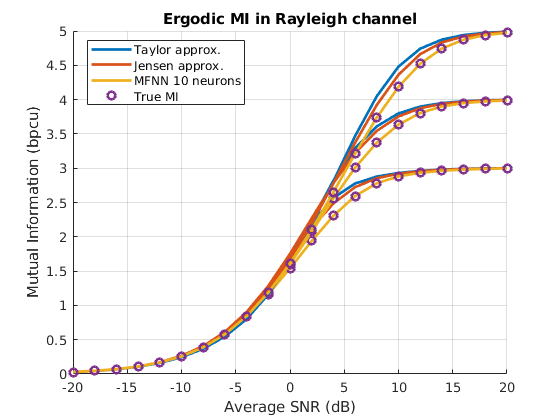}
	\caption{Ergodic MI obtained after averaging the instantaneous MI calculated with each method for a Rayleigh channel (QPSK, 8PSK and 16QAM).}
	\label{fig:ergodic}
\end{figure}

Finally,  the ergodic MI of an $2\times2$ Index Modulation system with QPSK, 8PSK and 16QAM constellations under Rayleigh fading is shown in Fig. \ref{fig:ergodic}. For each value of SNR, $100$ realizations of the channel matrix are generated, similarly to the NN dataset, and the true MI in each case is calculated with a Monte Carlo simulation with $1,000$ realizations of the noise. Afterwards, the ergodic MI for each SNR point is calculated by averaging the instantaneous values of the MI. The true ergodic MI, shown with circles, is compared with that obtained by averaging the instantaneous MI calculated with each method, the two analytical approximations and the $10$ neurons neural network. As it can be seen, the neural network matches perfectly the true ergodic MI, which is overestimated by the other methods for moderate values of the SNR. 

In addition to the estimation accuracy, the computational complexity is key for practical implementation: the MI computation must be done at the receive side, which has knowledge of the SNR $\gamma$ and the channel matrix $\H$. This evaluation must be such that an on-line tracking of the channel is made to report the estimated MI back to the receiver.

In this regard, Table \ref{tab:comparison-time} shows the computational complexity of each method after counting the number of mathematical operations required to compute the MIs for the three constellations. In the case of the analytical approximations, the table shows a lower bound of the number of operations since it only counts the most demanding instructions, which are repeated $(2M)^2$ times. In the case of the NN, all the required operations are counted, including the preprocessing of $\gamma$ and $\H$ to calculate the inputs of the NN. 

The numbers in Table \ref{tab:comparison-time} reveal that the MFNN is not only more accurate, but also less computationally demanding. The MFNN requires about $90$ times fewer multiplications and non linear operations than the analytical approximations. This is in line with the required time to compute with Matlab\textsuperscript{\tiny\textregistered} the three MIs for all the testing dataset, in a computer equipped with an i7-4510U 2 GHz processor. From another point of view,  the laptop only allows to make an estimation of the MIs every $5.5$ ms with the Taylor approximations, which gets reduced to $0.1$ ms with the NN. With respect to the off-line training duration, each training of the NN took typically less than $3$ minutes.

\begin{table*}[]
	\centering
	\begin{tabular}{|l|c|c|c|}
		\hline
		&\textbf{ Taylor approximation} & \textbf{Jensen based approximation} &\textbf{ MFNN option (v) 20 neurons} \\ \hline
		Real products & $7,168$ & $32,800$ & $368$ \\ \hline
		$\exp(\cdot)$ & $672$ & $1,344$ & $20$ \\ \hline
		$\log_2(\cdot)$ & $112$ & 3 & - \\ \hline
		Other non-linear operations & $1,344$ & - & $3$ \\ \hline
		\textbf{Time for $\mathbf{7,500}$ calculations} & $\mathbf{41}$ \textbf{s} & $\mathbf{76}$ \textbf{s} & $\mathbf{0.80}$ \textbf{s} \\ \hline	
	\end{tabular}
	\caption{Comparison of complexity and computational time of the MFNN and the two methods of the literature. Total number of operations for computing the three MI values (QPSK, 8PSK and 16QAM) are given, as well as the computational time required for calculating $7,500$ values of these MIs with Matlab\textsuperscript{\tiny\textregistered} running in a laptop.}
	\label{tab:comparison-time}
\end{table*}

\section{Extension to a higher number of antennas}
\label{sec:extension}

This paper is focused mainly on a SM system with two dimensions in order to study thoroughly the impact of the different input features selection. However, this approach can be easily extended to compute the MI in scenarios with a higher number of antennas. This section aims to explain how to apply the same philosophy to obtain the MI of a $4\times4$ and $8\times8$ SM systems, providing also some cles to keep the complexity bounded for higher numbers of antennas. 

Let us recall Table \ref{tab:features}, which portraits several selections of the NN input features for the case of $2$ antennas. As can be seen, the performance improves from the top to the botton; in order to avoid too a large number of input features in systems with a higher number of antennas, we propose to apply option (ii) as a trade-off between performance and complexity. This option makes use of the channel column norms (scaled by the SNR) and the angles. 
Whilst the number of norms increases only linearly with the transmit antennas, the number of angles raises rapidly with $N_t$ since there are $2\binom{N_t}{2}$ angles, a tuple ($\Theta_H,\varphi$) for each possible combination of two transmit antennas. For example, in a system with $16$ antennas there are $120$ pairs of angles. However, we have found out that it is not necessary to give the values of all the angles explicitly to the neural network. A few values characterizing  the statistical distribution of the angles suffice for the NN to  estimate the MI with an MSE similar to those values reported in Table \ref{tab:mse-NNs}.

The MI evaluation in  a SM system with $4\times4$ antennas can be easily done with an MFNN trained with the proper dataset, obtained now with $4\times4$ Rayleigh matrices, and using as input features the four values of $\gamma\lVert \h_l \rVert^2$ and the six pairs of angles $(\theta_H,\varphi)$. In the case of an $8\times8$ IM system we propose to reduce the $\binom{8}{2}=28$ pairs of angles to just $Q$ values per type of angle (Hermitian and Kasner). These $Q$ values are the quantiles of the angles distribution for $Q$ probabilities taken from $0$ to $1$ at equal steps. For example, for $Q=5$ the distribution of the angles is characterized by the minimum, the $25$th percentile, the $50$th percentile (the median), the $75$th percentile and the maximum. Therefore, the MFNN for obtaining the set of MI of an $8\times8$ IM system has as input features the $8$ columns norms $\gamma \lVert \h_l \rVert^2$, and the $Q$ quantiles of the Hermitian angle $\Theta_H$ and the Kasner's pseudoangle $\varphi$, respectively.

For testing purposes, we have generated two additional datasets, one with $50,000$ $4\times4$ Rayleigh matrices and another with $25,000$ $8\times8$ Rayleigh matrices. Again, each channel matrix has associated a random SNR value between $-20$ and $20$ dB, and we have calculated the MI of each pair ($\gamma,\H$) for several constellations (QPSK, 8PSK and 16QAM) using Monte Carlo simulations. Following the same procedure of training and testing described in Section \ref{sec:results}, we have obtained two trained MFNNs for calculating the three MI values of $4\times4$ and $8\times8$ SM systems, respectively. 

Table \ref{tab:4x4-8x8} sums up the results obtained with these two neural networks. Note that for the $8\times8$ system we have used only $Q=5$ quantiles. If we compare the results of $4$ and $8$ antennas provided in Table \ref{tab:4x4-8x8} with  those obtained with the best network for the $2$ antennas scenario from Table \ref{tab:comparison}, using $20$ neurons and input features option (v), the NN performs a little worse in the setup with more antennas, although the errors are of the same magnitude. However, a fair comparison with the NN for $2\times2$ SM which uses the same type of input features, i.e., option (ii) from Table \ref{tab:mse-NNs}, reveals that the MSE is slightly better when the number of dimensions grows.

\begin{table*}[]
 	\centering
 	\begin{tabular}{|l|c|c||c|c||c|c||c|c|}
 		\hline
 		\multirow{2}{*}{} & \multirow{2}{*}{\textbf{NN input features}} & \multirow{2}{*}{\textbf{Global MSE}} & \multicolumn{2}{c||}{\textbf{QPSK}} & \multicolumn{2}{c||}{\textbf{8PSK}} & \multicolumn{2}{c|}{\textbf{16QAM}} \\ \cline{4-9} 
 		&  & & $\mathbf{3\sigma}$ & \textbf{Max. error} & $\mathbf{3\sigma}$ & \textbf{Max. error} &  $\mathbf{3\sigma}$ & \textbf{Max. error} \\ \hline
 		\textbf{SM} $\mathbf{2\times2}$ \textbf{(20 neurons), option (ii)} & $2 + 2\times 1 = 4$ & $3.36 \cdot 10^{-4}$  & $0.078$  & $0.436$   & $0.041$ & $0.342$ & $0.034$  & $0.322$  \\ \hline
 		\textbf{SM} $\mathbf{4\times4}$ \textbf{(20 neurons)} & $4 + 2\times6 = 16$ & $2.40 \cdot 10^{-4}$  & $0.047$  & $0.169$   & $0.050$ & $0.200$ & $0.041$  & $0.137$  \\ \hline
 		\textbf{SM} $\mathbf{8\times8}$ ($\mathbf{Q=5}$, \textbf{20 neurons)} & $8 + 2\times5 = 18 $ & $5.06 \cdot 10^{-5}$   & $0.022$   & $0.050$   & $0.023$  & $0.061$ & $0.018$  & $0.046$  \\ \hline
 	\end{tabular}
 	\caption{Performance of the MFNNs for obtaining the MI for QPSK, 8PSK and 16QAM of a SM system with $2$, $4$ and $8$ transmit and receive antennas.}
 	\label{tab:4x4-8x8}
\end{table*}

\section{Conclusions}
\label{sec:conclusions}
The implementation of next generation adaptive Spatial Modulation links requires practical mechanisms to estimate the mutual information for a given signalling strategy.  An accurate and timely  computation  of this constrained capacity serves to adapt the  constellation order, and apply a fine tuning of the coding rate of the channel encoder, providing a better fit to the maximum achievable rate. The method proposed in this paper to calculate the constrained capacity is a very simple Multilayer Feedforward Neural Network, which can obtain the Mutual Information for different symbol constellations simultaneously. 
The neural network is more accurate and less computationally demanding than the analytical approximations existing in the literature. 
 
 \section{Acknowledgments}
This work was funded by the Xunta de Galicia (Secretaria Xeral de Universidades) under a predoctoral scholarship (cofunded by the European Social Fund) and it was partially funded by the Agencia Estatal de Investigaci\'on (Spain) and the European Regional Development Fund (ERDF) under project MYRADA (TEC2016-75103-C2-2-R). It was also funded by the Xunta de Galicia and the ERDF (Agrupaci\'on Estrat\'exica Consolidada de Galicia accreditation 2016-2019). Furthermore, this work has received funding from the Spanish Agencia Estatal de Investigaci\'on under project TERESATEC2017-90093-C3-1-R (AEI/FEDER,UE); and from the Catalan Government (2017 SGR 891 and 2017 SGR 1479).

\bibliographystyle{unsrt}
\bibliography{Bibliografia_new}
\end{document}